\documentclass[twocolumn,a4paper,superscriptaddress,floatfix,rmp]{revtex4}

\usepackage{mathbbol,amssymb,latexsym,amsfonts,amsmath,amsthm}
\usepackage{graphicx,color}
\usepackage{natbib}
\usepackage{mathpazo}

\newcommand{\ppost}{P_{\mathrm{post}}}
\newcommand{\A}{\mathcal{A}}
\newcommand{\T}{\mathcal{T}}
\renewcommand{\L}{\mathcal{L}}
\newcommand{\C}{\mathcal{C}}
\newcommand{\D}{\mathcal{D}}
\newcommand{\E}{\mathcal{E}}
\newcommand{\Rr}{\mathcal{R}}

\begin{document}

\title[Module networks revisited]{Module networks revisited:
  computational assessment and prioritization of model predictions}

\author{Anagha Joshi}

\affiliation{Department of Plant Systems Biology, VIB, Technologiepark
  927, B-9052 Gent, Belgium}

\affiliation{Department of Molecular Genetics, UGent, Technologiepark
  927, B-9052 Gent, Belgium}

\author{Riet De Smet}

\affiliation{CMPG, Department of Microbial and Molecular Systems,
  KULeuven, Kasteelpark Arenberg 20, B-3001 Leuven, Belgium}

\author{Kathleen Marchal}

\affiliation{CMPG, Department of Microbial and Molecular Systems,
  KULeuven, Kasteelpark Arenberg 20, B-3001 Leuven, Belgium}

\affiliation{ESAT-SCD, KULeuven, Kasteelpark Arenberg 10, B-3001
  Leuven, Belgium}

\author{Yves Van de Peer}

\author{Tom Michoel}

\email[Corresponding author, E-mail: ]{tom.michoel@psb.ugent.be}

\affiliation{Department of Plant Systems Biology, VIB, Technologiepark
  927, B-9052 Gent, Belgium}

\affiliation{Department of Molecular Genetics, UGent, Technologiepark
  927, B-9052 Gent, Belgium}

\begin{abstract}
  \textbf{Motivation:} The solution of high-dimensional inference and
  prediction problems in computational biology is almost always a
  compromise between mathematical theory and practical constraints
  such as limited computational resources. As time progresses,
  computational power increases but well-established inference methods
  often remain locked in their initial suboptimal solution.

  \textbf{Results:} We revisit the approach of Segal \textit{et~al.}
  (2003) to infer regulatory modules and their condition-specific
  regulators from gene expression data. In contrast to their direct
  optimization-based solution we use a more representative
  centroid-like solution extracted from an ensemble of possible
  statistical models to explain the data. The ensemble method
  automatically selects a subset of most informative genes and builds
  a quantitatively better model for them.  Genes which cluster
  together in the majority of models produce functionally more
  coherent modules. Regulators which are consistently assigned to a
  module are more often supported by literature, but a single model
  always contains many regulator assignments not supported by the
  ensemble.  Reliably detecting condition-specific or combinatorial
  regulation is particularly hard in a single optimum but can be
  achieved using ensemble averaging.

  \textbf{Availability:} All software developed for this study is
  available from
  \url{http://bioinformatics.psb.ugent.be/software/}.
  
  \textbf{Supplementary information:} Supplementary data and figures
  are available from
  \url{http://bionformatics.psb.ugent.be/supplementary_data/anjos/module_nets_yeast/}.
\end{abstract}

\maketitle

\section{Introduction}
\label{sec:intro}

One of the central goals of the top-down approach to systems biology
is to infer predictive mathematical network models from
high-throughput data.  Much of the driving force for the development
of network inference methods has come from the availability of various
types of large-scale data sets for particular model organisms like
\textit{S.~cerevisiae} and \textit{E. coli}.  In contrast, data
generation for other organisms has been much slower and mainly focused
on gene expression data.  These gene expression data sets for
typically more complex organisms pose their own challenges, such as a
higher number of genes, limited number of experimental conditions, and
supposedly a more complex underlying transcriptional network.
Therefore, improvement and refinement of methods for network inference
from gene expression data continues to be of great interest.  Several
reviews on a variety of methods have been written
\citep{friedman2004,gardner2005,bansal2007,bussemaker2007}, and
development of new methods remains an active area of research
\citep{basso2005,faith2007,bonneau2006,alter2005}.  Here we revisit
the module network method of \cite{segal2003} to infer regulatory
modules and their condition-specific regulators from gene expression
data and show that better and more refined module networks can be
obtained by using advanced statistical and computational methods.
These improvements concern the use of Monte Carlo \citep{liu2004} and
ensemble strategies \citep{carvalho2007,webb-robertson2008}.

Following \cite{hartwell1999} a `module' is to be viewed as a discrete
entity composed of many types of molecules and whose function is
separable from that of other modules. Understanding the general
principles that determine the structure and function of modules and
the parts they are composed of can be considered one of the main
problems of contemporary systems biology \citep{hartwell1999}. The
module network method of \cite{segal2003} addresses this problem using
gene expression data as its input.  It has yielded novel biological
insights in a number of complex eukaryotic systems
\citep{segal2003,lee2006,segal2007,zhu2007b,li2007,novershtern2008}
and has been the source of inspiration for numerous computational
approaches to network inference as evidenced by its high number of
citations.  A module network is a probabilistic graphical model
\citep{friedman2004} which consists of modules of coregulated genes
and their regulatory programs. A regulatory program uses the
expression level of a set of regulators to predict the condition
dependent mean expression of the genes in a module. \cite{segal2003}
used a deterministic optimization algorithm that searches
simultaneously for a partition of genes into modules and a regulation
program for each module. We consider both as separate tasks.  When
searching for modules, often many local optima exist with partially
overlapping modules differing from each other in a few genes.  We use
a Gibbs sampling approach for two-way clustering of genes and
conditions to generate an ensemble of partially overlapping partitions
of genes into modules and produce an ensemble averaged solution
\citep{joshi2007}.  This centroid solution consists of so-called
\emph{tight clusters}, subsets of genes which consistently cluster
together in almost all local optima.  We also use a probabilistic
method for learning regulatory programs. These regulatory programs
take the form of fuzzy decision trees with regulator expression levels
at the decision nodes and generalize the regression tree approach of
\cite{segal2003}.  By summing the strength with which a regulator
participates in each member of an ensemble of regulatory programs for
a certain module, we obtain a regulator score which gives a
statistical confidence measure for the assignment of that regulator.
Together, the Gibbs sampling cluster algorithm and probabilistic
regulatory program learning provide a computationally efficient method
to generate ensembles of module networks from which a centroid-like
summarization can be constructed.

We have applied this ensemble method to the very same data set as
\cite{segal2003} and performed several comparison tasks.  First, we
considered the probabilistic models and evaluated them on training as
well as test data. We show that the model inferred by \cite{segal2003}
is equivalent to a single instance of the ensemble of models inferred
by our algorithm. The tight clusters obtained from the ensemble
solution generate a quantitatively better model than each of the
single instances, including the model of \cite{segal2003}. Second, we
compared the clustering of genes.  Tight clusters are in general more
functionally coherent and improve the original modules in two ways.
They can remove spurious profiles and fetch only the core of tightly
coexpressed genes from a single module, or they can merge separate but
related modules into one cluster. Third, we used the regulator score
to analyze the network of modules and their associated regulators from
\cite{segal2003}. We show that this network contains both high- and
low-scoring regulators and that several high-scoring regulators are
missed by the solution of \cite{segal2003}. In general, regulator
assignments which can be validated by external sources such as ChIP
data or literature are highly ranked.  In combination with the tight
clusters, the probabilistic method assigns more regulators supported
by literature and the clusters to which they are assigned contain a
higher ratio of known targets compared to the module network of
\cite{segal2003}.  Fourth, we show that the regulator scoring scheme
can also be used to infer context-specific and combinatorial
regulation by identifying pairs of regulators which occur
significantly often together in the same regulation program.

Finally we have applied the ensemble method to a bHLH module network
that was recently inferred for mouse brain \citep{li2007}.
\cite{li2007} used their module network to make several hypotheses
about modes of combinatorial regulation among different brain tissues.
We show that only few of these hypotheses are statistically supported
by the ensemble method.  This example illustrates the usefulness of an
approach which can generate internal significance measures, in
particular if no other data sources are available to validate
hypotheses generated by a single local optimum.

Together all these results convincingly show that the ensemble method
for learning module networks significantly improves the direct
optimization method of \cite{segal2003}. Unlike a single optimum,
ensemble averaging allows the assessment and prioritization of the
statistically most reliable modules and their condition-specific
regulators.  Such high-confidence modules can be used directly for
generating experimentally verifiable hypotheses or can be integrated
with other, perhaps smaller-scale, data sources to create a more
comprehensive view of the underlying networks.

\section{Results and discussion}
\label{sec:results}

\subsection{Data and procedure}
\label{sec:data-procedure}

We obtained all data from the supplemental website of
\cite{segal2003}, including expression data, gene modules and
regulatory programs.  Using the Gibbs sampler we generated 12
different partitions of genes into modules which were combined into
one set of tight clusters.  The number of clusters is determined
automatically by the Gibbs sampler and ranges from 65 to 78 in the
different runs, compared to the predefined value of 50 of
\cite{segal2003}. 1892 of the 2355 genes in the data set could be
assigned with high confidence to 69 tight clusters.  To generate
regulator assignment scores, we learned 10 probabilistic regulation
programs per module with 100 regulator and split value pairs sampled
per regulation program node. More details about these procedures are
given in the Methods. This resulted in four different module network
models:
\begin{enumerate}
\item SCSR: Segal clusters with Segal regulation programs,
  corresponding exactly to the results of \cite{segal2003}.
\item SCPR: Segal clusters with probabilistic regulation programs.
\item GCPR: Gibbs sampler clusters (single run) with probabilistic
  regulation programs.
\item TCPR: Tight clusters (multiple Gibbs sampler runs combined) with
  probabilistic regulation programs.
\end{enumerate}

\subsection{Model evaluation}
\label{sec:model-eval}

A module network infers a probabilistic model which explains relations
between expression levels of a set of genes. More precisely, there is
a probability distribution $p(x_1,\dots,x_N)$ which computes the
probability (density) to observe a particular combination of
expression levels $x_i$ for a set of $N$ genes. This probabilistic
model predicts the response in expression of genes in a module upon
perturbations of its regulators, such as knock-out or overexpression,
and thus yields biologically verifiable hypotheses.  For a module
network, the distribution $p(x_1,\dots,x_N)$ is a product of $N$
factors (see Methods), so we consider the normalized quantity
$\L=\frac1N \log p(x_1,\dots,x_N)$ which can be compared between
models with potentially different numbers of genes.  Higher values of
$\L$ mean better explanation of the data by the model, \textit{i.e.}
more accurate prediction of the outcome of new experiments.

\begin{figure}
  \centering
  \includegraphics[width=\linewidth]{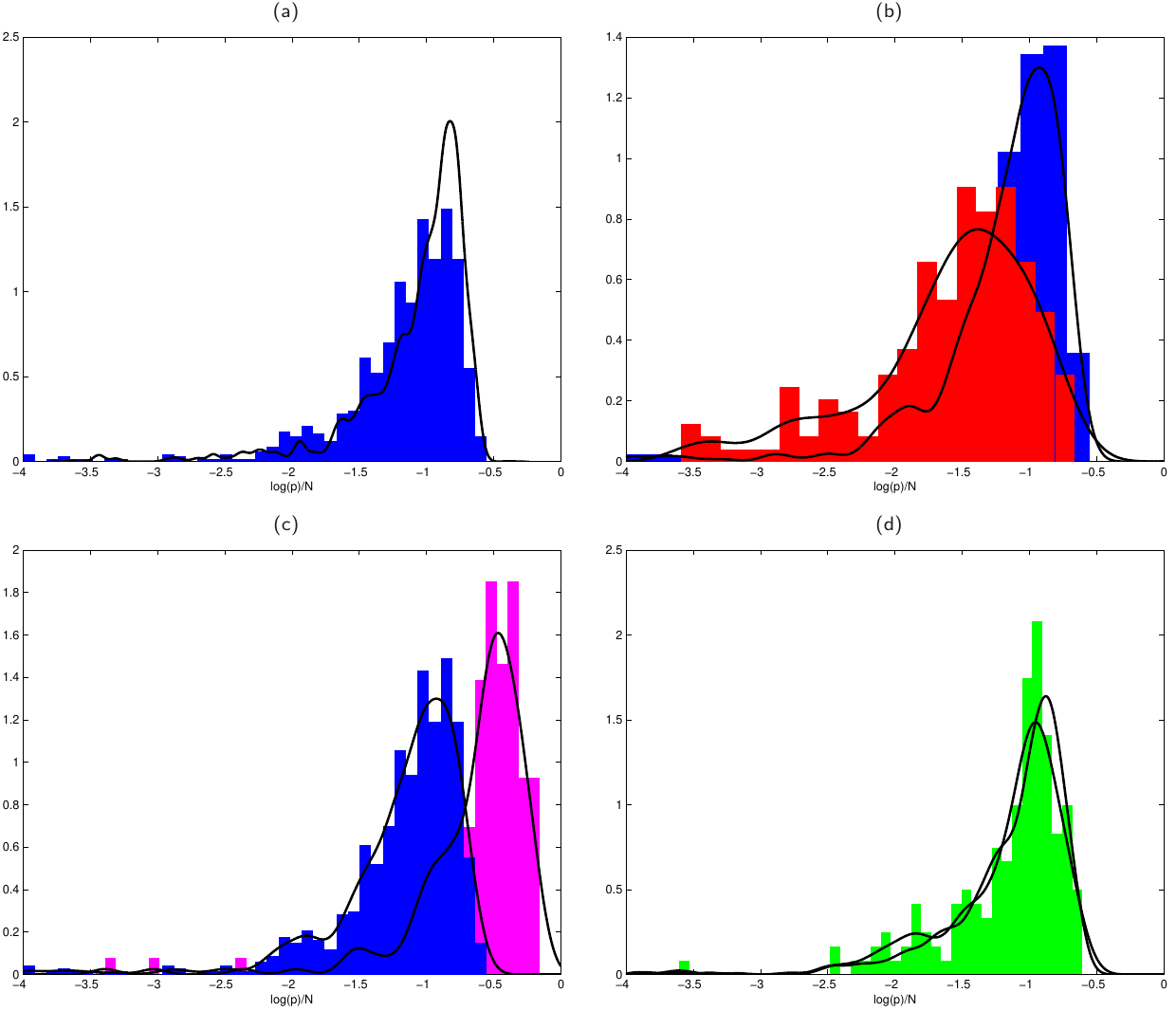}
  \caption{Model evaluation experiments. \textbf{(a)} Histogram of
    $\L=\frac 1N\log p(x_1,\dots,x_N)$ for SCPR (blue) and
    non-parametric fit of the histogram for GCPR (black curve).
    \textbf{(b)} Histogram of $\L$ for SCSR (red) overlayed on
    histogram of $\L$ for SCPR (blue), with non-parametric fits (black
    curves). \textbf{(c)} Histogram of $\L$ for SCPR (blue) overlayed
    on histogram of $\L$ for TCPR (magenta), with non-parametric fits
    (black curves). \textbf{(d)} Histogram and non-parametric fit
    (left black curve) of $\L$ for GCPR learned on training data and
    evaluated on test data (green) and non-parametric fit of the same
    models evaluated on training data (right black curve).  All
    histograms and curves are normalized to have area equal to 1.}
  \label{fig:eval}
\end{figure}

First we performed evaluations on each of the conditions in the
original data set. Figure \ref{fig:eval} (a) shows that the histogram
of $\L$-values for SCPR fits well within a non-parametric curve fit of
the histogram for GCPR. This implies that the clusters found by
\cite{segal2003} are equivalent to one local optimum identified by the
Gibbs sampler procedure. Figure \ref{fig:eval} (b) shows the histogram
of $\L$-values for SCSR (red) overlayed on the histogram for SCPR
(blue), both with non-parametric curve fits. The mean $\L$-values
obtained by SCPR are higher than SCSR by a one-tailed $t$-test
($\alpha = 0.01$) proving that probabilistic regulation programs give
a better explanation of the data. In Figure \ref{fig:eval} (c) we
compared SCPR to TCPR.  TCPR has a higher mean $\L$ than SCPR with a
one-tailed t-test ($\alpha = 0.01$). This shows that tight clusters
are selecting a subset of genes which are the most informative and
therefore generate a better model.

Next we tested how well these models explain unseen data by performing a
cross-validation experiment. We removed $10\%$ of the conditions at
random from the complete data (the test set) and ran the Gibbs sampler
once on the remaining $90\%$ (the training set). The resulting model
was then evaluated on the test set. This procedure was repeated 10 times
and all test set evaluation values were collected in one histogram and
compared to the training set values (Figure \ref{fig:eval} (d)).  The
curve of the test set is slightly shifted to the left with respect to
the training set curve, as one would expect, but both curves have the
same mean with a one-tailed $t$-test ($\alpha = 0.01$). This shows
that the probabilistic models indeed generalize to unseen data.

\subsection{Gene clustering improvement}
\label{sec:gene-clust}

\begin{figure}[!tpb]
  \centering
   \includegraphics[width=\linewidth]{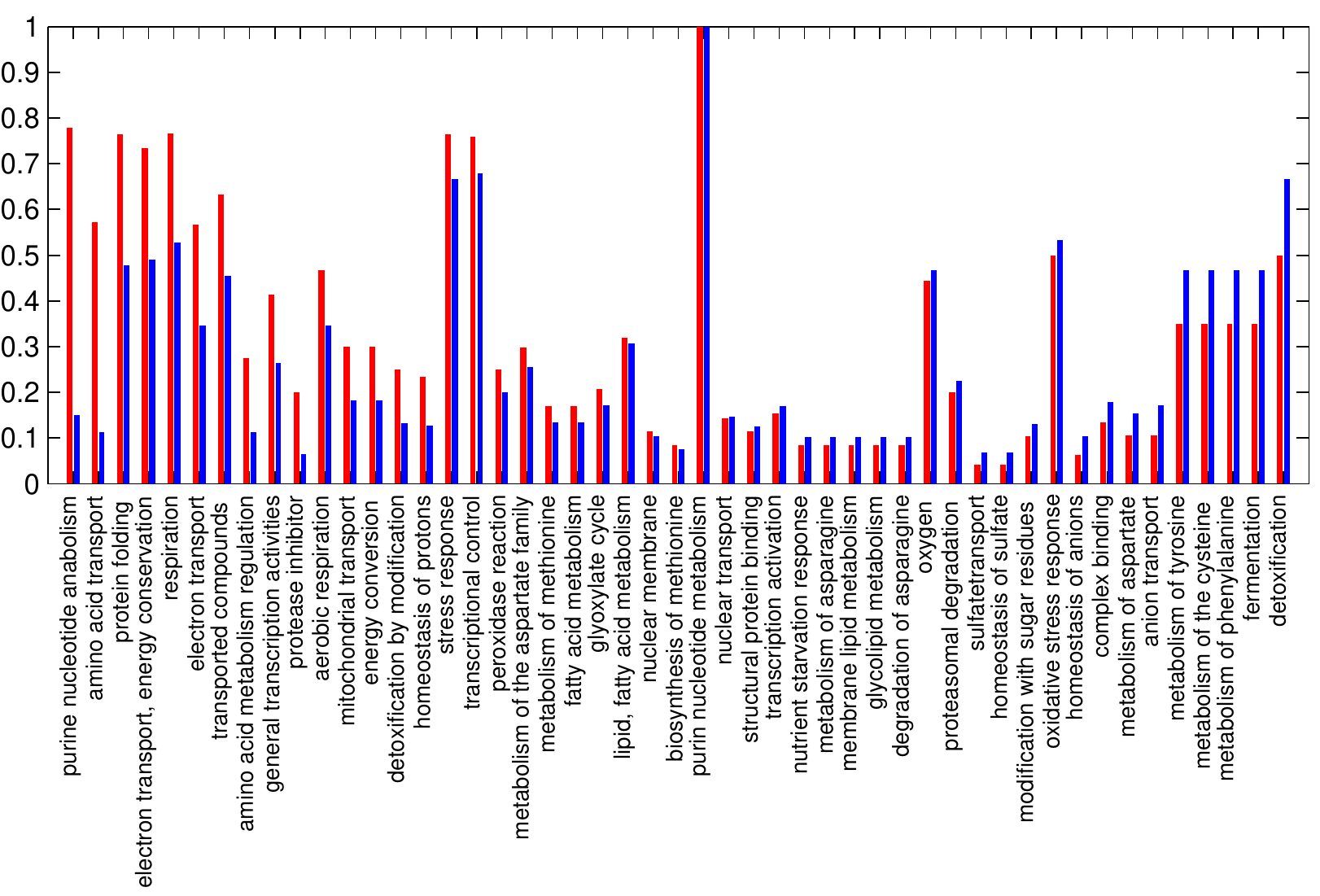}
   \caption{Histogram of the highest fraction of genes in one module
     in a MIPS functional category for TC (red) and SC (blue), sorted
     by ratio difference.}
  \label{fig:mips}
\end{figure}

We have shown in the previous section that SCPR is equivalent to GCPR
but TCPR gives a better model over SCPR. We also observe that tight
clusters (TC) are overall more functionally coherent than the clusters
obtained in \cite{segal2003} (SC).  Figure \ref{fig:mips} shows the
fraction of genes in a cluster belonging to a MIPS functional category
which is significantly overrepresented ($p<0.001$) in SC and TC.
Several examples illustrate the general trend seen in this figure. In
TC-40, 4/7 genes are involved in amino acid transport compared to
SC-27 with 8/53 genes. In TC-27, 7/9 genes belong to purine nucleotide
anabolism compared to SC-11 with 6/53 genes.

\begin{figure}[!tpb]
  \centering
   \includegraphics[width=\linewidth]{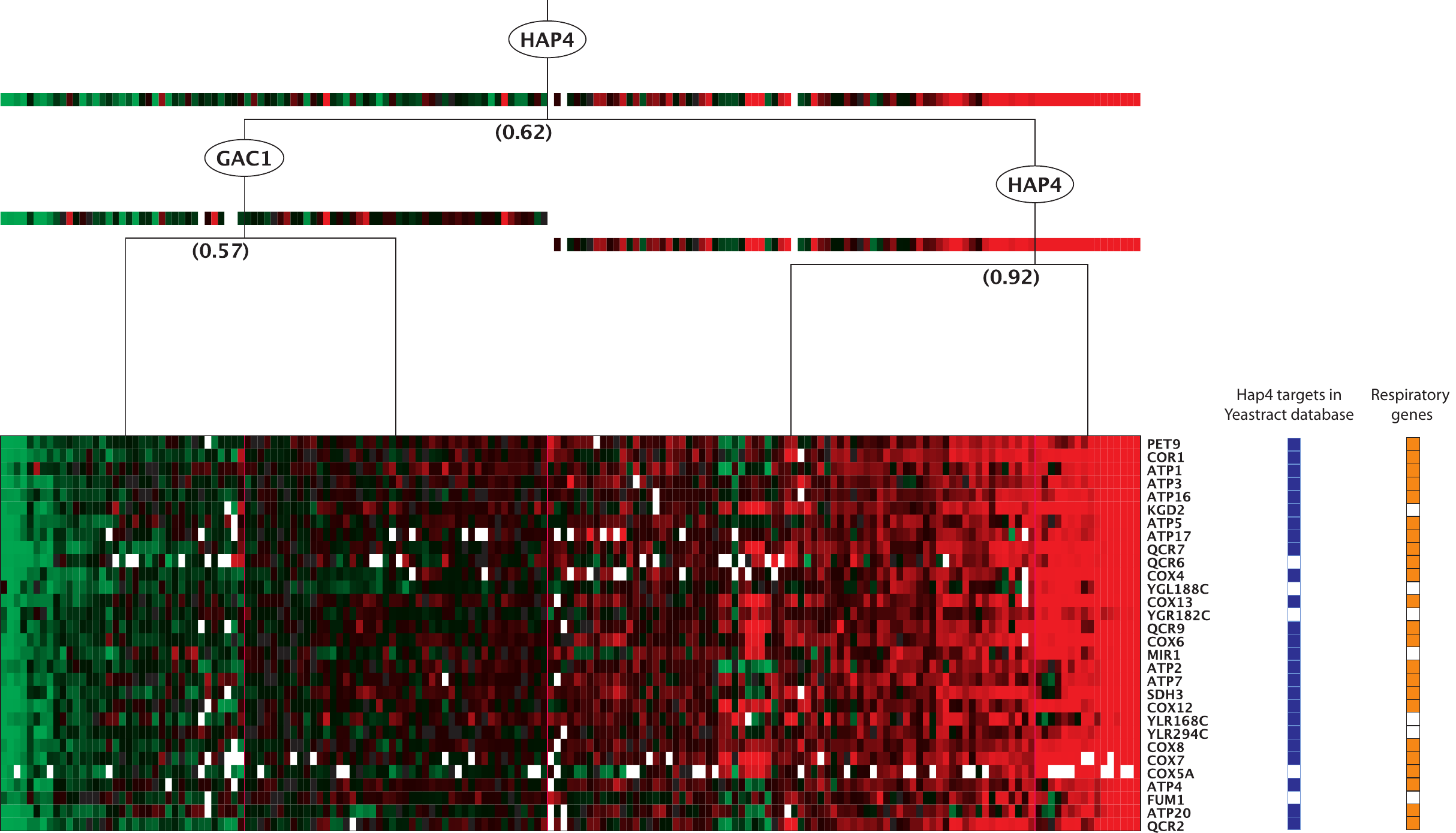}
   \caption{TC-7 with Hap4 assigned as a top regulator.  Genes known
     to be regulated by Hap4 in YEASTRACT are marked in blue and those
     involved in respiration are marked in orange.}
  \label{fig:hap4}
\end{figure}

Segal cluster 1 (SC-1) contains 55 genes, out of these 32 ($58\%$) are
validated targets of Hap4, a global regulator of respiratory genes,
according to the YEASTRACT database \citep{teixeira2006}.  This
cluster has maximum overlap with tight cluster 7 (TC-7) with 30 genes
out of which 25 ($83\%$) are known Hap4 targets.  The five remaining
genes are Qcr6, Cox5a and Fum1, all located in mitochondrion and
involved in respiration, and two unknown genes Ygl188c and Ygr182c.
With 24/30 respiratory genes ($80\%$), TC-7 even improves on COGRIM
\citep{chen2007} which combines multiple data sources.  Using
expression data alone (the same data set as \cite{segal2003}),
\cite{chen2007} obtain a cluster with 32/51 ($62\%$) genes belonging
to MIPS respiration category.  Using both ChIP and expression data,
they obtain a cluster with 23/34 ($68\%$) respiratory genes,
significantly lower than TC-7. Figure \ref{fig:hap4} shows TC-7 with
known Hap4 targets and respiratory genes marked in blue and orange
respectively.

\begin{figure}[!tpb]
  \centering
  \includegraphics[width=\linewidth]{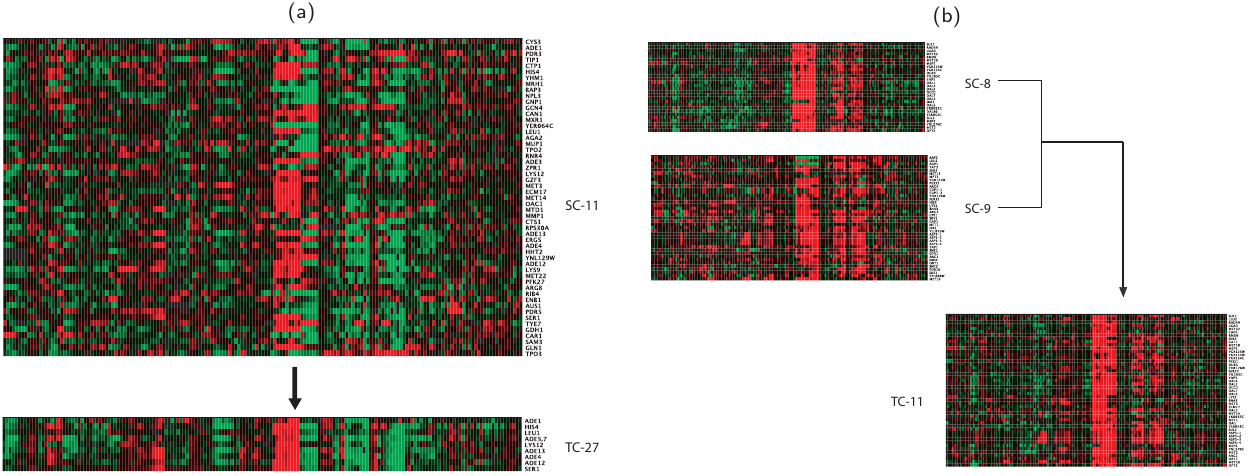}
  \caption{\textbf{(a)} TC-27 fetches the core of tightly coexpressed
    genes from SC-11; $67\%$ genes in TC-27 are known to be Bas1
    targets. \textbf{(b)} SC-8 and SC-9 which have similar expression
    are merged into TC-11. SC-8, SC-9 and TC-11 all are enriched for
    Gat1 targets.}
  \label{fig:break_make}
\end{figure}

TC-27 contains nine genes which form a subset of SC-11 containing 53
genes (Figure \ref{fig:break_make} (a)). Six genes ($67\%$) in this
cluster are known Bas1 targets compared to only $18\%$ Bas1 targets in
SC-11.  TC-28 and TC-37 contain $70\%$ and $100\%$ known targets of
Msn4.  These clusters have a large overlap with SC-3 and SC-41
respectively, which have $55\%$ and $93\%$ known targets of Msn4.
TC-1 consists of 51 genes, out of which 28 ($55\%$) are known to be
Swi4 targets.  This module merges genes from SC-10, 29 and 30.  They
have 4/37 ($11\%$), 19/41 ($46\%$) and 8/30 ($27\%$) Swi4 targets
respectively.  TC-11 contains genes of SC-8 and SC-9 whose highest
ranked regulator is Gat1 (see Section \ref{sec:regul-assignm}) (Figure
\ref{fig:break_make} (b)).  YEASTRACT data confirms $17\%$ of these
targets, while for SC-8 and 9 overall $15\%$ targets are confirmed by
YEASTRACT.  TC-35 is overrepresented for genes involved in RNA export
from nucleus ($p$-value $10^{-8}$).  It overlaps with SC-19, 31 and 36
($p$-values $\sim 10^{-3}$).  TC-31 contains genes mainly involved in
ribosomal biogenesis ($p$-value $10^{-13}$) and combines relevant
genes from SC-13, 14 and 15 ($p$-values $\sim 10^{-4}$).

We conclude that tight clusters improve clustering results obtained by
\cite{segal2003} in two ways. They can fetch only the core of tightly
coexpressed genes from a SC (Figure \ref{fig:break_make} (a)), or they
can merge clusters which were separate in SC (Figure
\ref{fig:break_make} (b)).

\subsection{Regulator assignment prioritization}
\label{sec:regul-assignm}

The ensemble approach generates multiple equally plausible regulatory
programs for a single module in a probabilistic fashion.  The
regulator assignment score which takes into account how often a
regulator is assigned to a module, with what score, and at which level
in the regulation tree, can therefore be used to prioritize regulators
(highest regulator score gets topmost rank).  

\begin{figure*}[!tpb]
  \centering
   \includegraphics[width=\linewidth]{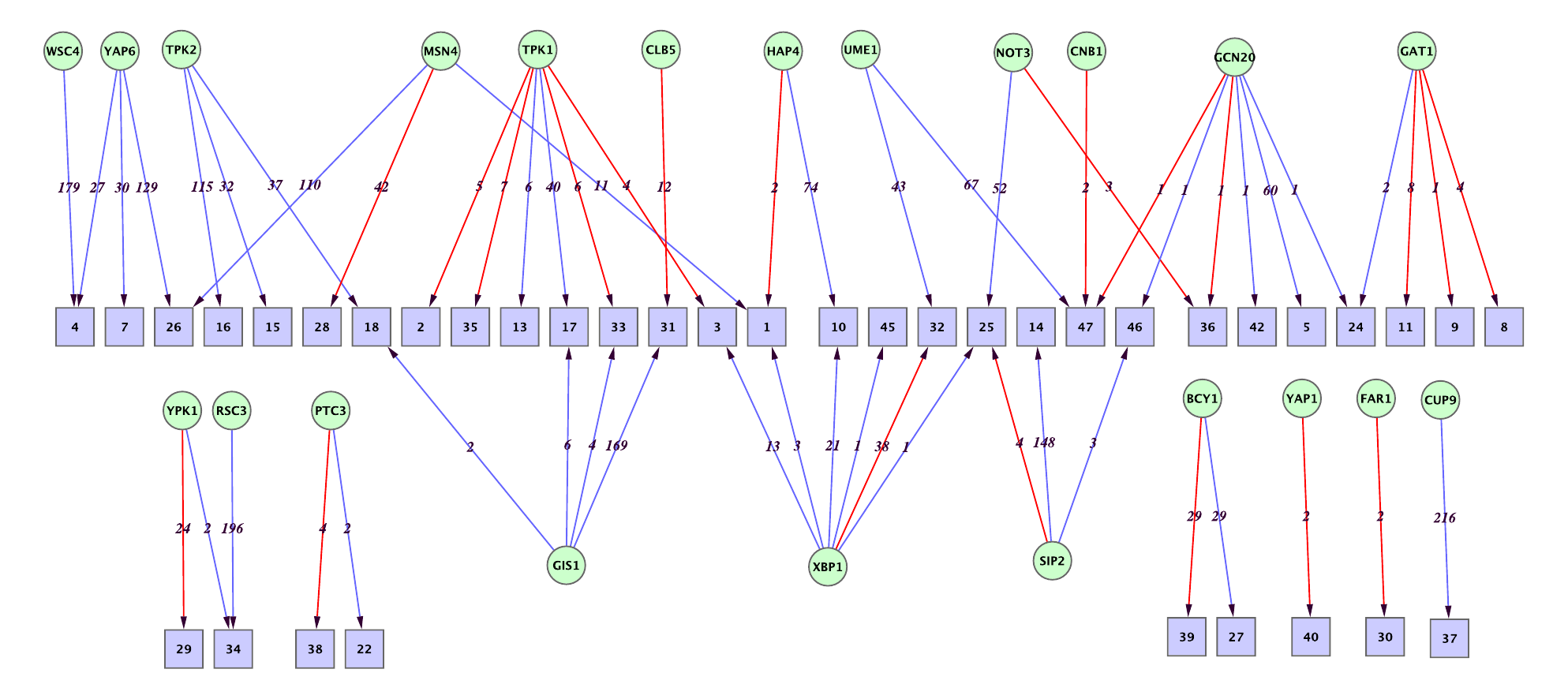}
   \caption{Module network inferred by \cite{segal2003} with
     edge-ranks computed by the ensemble method described in the
     current paper. Red edges mean the
     module is overrepresented in known targets of the connected
     regulator.}
  \label{fig:regs}
\end{figure*}

First we consider only the difference between probabilistic regulator
assignment and the original method by comparing SCSR with SCPR, hence
keeping the gene modules the same for both methods.  Figure
\ref{fig:regs} shows regulator-module links in SCSR (cfr. Figure 5 in
\cite{segal2003}). The edges colored red are the ones supported by
literature (data from \cite{segal2003}). To each edge we add the rank
with which it is assigned in SCPR.  Regulator-module links supported
by literature have often a higher rank. SCSR assigns Hap4, a global
regulator of respiratory genes, to SC-1. This cluster contains $58\%$
known Hap4 targets and Hap4 has second highest rank in SCPR. SCSR also
assigns Hap4 to SC-10 which contains genes involved in amino acid
metabolism. SC-10 has only 2/37 ($5\%$) known Hap4 targets according
to YEASTRACT and this assignment is ranked very low (rank 73) in SCPR.
Several high-ranking SCPR assignments which were missed by SCSR could
also be validated using \cite{harbison2004} data ($p$-value $<0.005$).
We assign Gal80, a transcriptional regulator involved in the
repression of Gal genes in the absence of glucose, with second rank to
SC-6. This is a cluster of four Gal genes, Gal1, Gal2, Gal7 and Gal10.
Met32, a zinc-finger DNA-binding protein involved in transcriptional
regulation of the methionine biosynthetic genes assigned with third
rank to SC-8, and Gis1, a histone demethylase assigned to SC-3 with
5th rank, are supported by YEASTRACT (respectively 5/29 and 6/31 known
targets).

\begin{figure}[ht!]
  \centering
   \includegraphics[width=\linewidth]{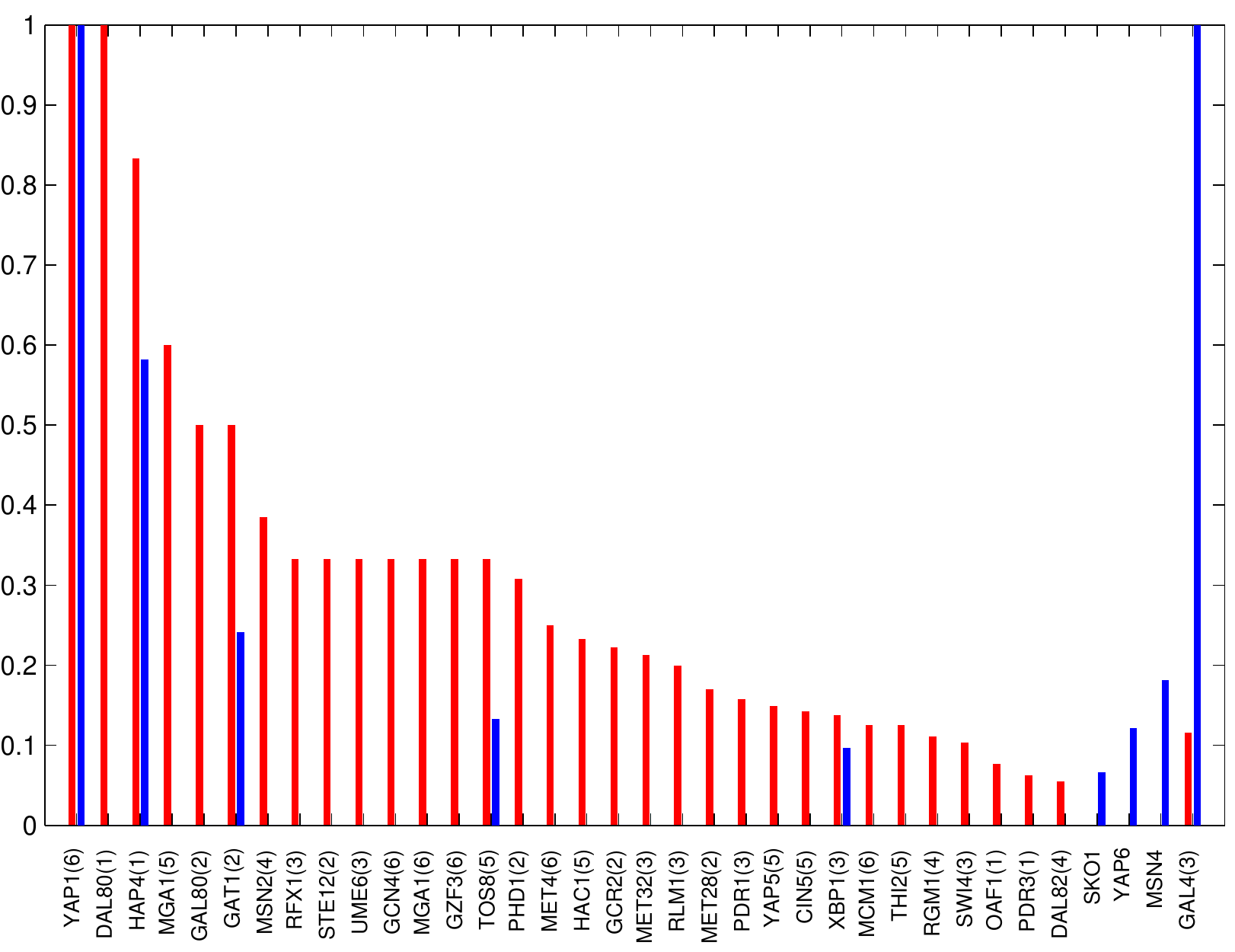}
   \caption{Histogram of the highest fraction of known targets of
     transcription factors in a module using TCPR (red) and SCSR
     (blue) according to the YEASTRACT database. The rank with which a
     regulator is assigned to a module in TCPR is indicated in
     brackets.}
  \label{fig:mips_regs}
\end{figure}

Next we compared TCPR with SCSR to analyse the combined improvement
made by ensemble averaging at the level of gene clustering as well as
at the level of regulator assignment.  For TCPR, we selected the top
six regulators for each cluster. This rank cutoff was determined as
follows. We computed the significance for the overlap between each
tight cluster and each transcription factor target set using the
YEASTRACT database.  A reference module network was formed by keeping
all transcription factor - tight cluster edges below a certain
$p$-value cutoff. By comparison with this reference network we found
that a rank cutoff of six gives the best overall $F$-measure score at
different $p$-value cutoffs (see Supplementary information). A similar
analysis for SCSR shows that the $F$-measure for TCPR is consistently
higher (see Supplementary information). To compare TCPR and SCSR in
more detail, we identified for each regulator the cluster with the
highest fraction of known targets in YEASTRACT.  Likewise we find the
best cluster for each regulator in SCSR.  Figure \ref{fig:mips_regs}
shows that TCPR assigns more regulators supported by YEASTRACT and
also that the clusters contain a higher ratio of known targets. There
are six regulators assigned by both methods, four of which HAP1, GAT1,
TOS8 and XBP1 all are assigned to clusters more enriched in their
known targets in the TCPR solution.

\subsection{Context-specific and combinatorial regulation}
\label{sec:comb-regul}

\cite{segal2003} used a decision tree approach to model regulatory
programs because it can represent, at least in principle,
context-specific and combinatorial regulation.  In the ensemble
language, context-specificity means a regulator gets a high overall
score by being assigned consistently to a lower, non-root level in the
set of decision trees for a certain module. In SCPR, 59 regulator
assignments divided over 39 (out of 50) modules have a significant
score contribution (value $>100$) from a non-root level (see the
Methods for the decomposition of the score function over different
tree levels).  Combinatorial regulation means two (or more) regulators
are consistently assigned together at different levels in all decision
trees.  Although this form of combinatorial regulation may correspond
to genuine biological combinatorial regulation, we take a strictly
data driven definition here: combinatorial regulation in the decision
tree sense means the expression levels of both regulators are needed
\emph{together} to explain the expression level of the module (`AND'
regulation).  Alternatively, two (or more) high-scoring regulators may
achieve their high rank from the same decision tree level (usually the
root level).  In this case both regulators explain the module equally
well \emph{alone} (`OR' regulation). In SCPR, there are a total of 100
regulator assignments with significant score contribution from the
root level (OR regulation), which can be combined with the 59
assignments at level 1 for potential AND combinatorial regulation.

Only few of the significant AND combinatorial regulation pairs are
present in the single-optimum solution of SCSR (see edge ranks in
Figure \ref{fig:regs}). SC-47 has Gcn20 as the highest ranked
regulator at level 0 and Cnb1 at level 1, and both assignments are
supported by literature. SC-36 has two validated regulators Gcn20 and
Not3 ranked first and third respectively in SCPR, but the score of
Not3 is low and not deemed significant.  SC-4 is an example of OR
regulation wrongly assigned in SCSR. In SCSR, Ypl230w is assigned at
level 0 and Gac1 at level 1, but in SCPR both are assigned at level 0
with first and third rank respectively and no high-scoring regulator
is found at level 1.  Some of the AND combinatorial regulation pairs
in SCPR that were missed in SCSR can be validated by YEASTRACT.  SC-40
has Tos8 assigned at level 0 (overall rank 1) and Yap1 at level 1
(overall rank 2). Tos8 has 3/15 known targets in this module while
Yap1 has all known targets (15/15).  SC-26 has Gac1 at root level
(overall rank 1) and Mal13 at level 1 (overall rank 2).  Mal13 has two
known targets (out of six known) in SC-26.

Due to the high number of possible regulator combinations, identifying
statistically significant regulation of AND-type is an even more
complex problem than simple regulator assignment. These examples show
that also for this problem, the ensemble approach is well suited.

\subsection{Module network in mouse brain}

Recently, \cite{li2007} reconstructed a bHLH transcription factor
regulatory network in mouse brain by a direct application of the
method of \cite{segal2003}. They selected a small data set of 198
genes and 22 conditions, built a module network using 22 bHLH
transcription factors as candidate regulators and assigned 15
different regulators to 28 modules (denoted again by SC), out of which
12 ($43\%$) have at least two genes in the same GO category.  Based on
the co-occurence of regulators in the regulation programs of
individual modules, \cite{li2007} make hypotheses about different
modes of coregulation among brain tissues which are currently not
confirmed by other data sources. We applied the ensemble method on
this data set and got 17 tight clusters (denoted by TC), out of which
11 ($65\%$) have at least two genes in the same GO category.

Only 11/28 SC have a high-scoring regulator with a significant score
contribution from a non-root level, compared to 39/50 for yeast.
\cite{li2007} use the co-occurrence of Neurod6 and Hey2 in the SR
regulation programs of SC-10, 15 and 27 to predict a cross-repression
between Neurod6 and Hey2 with different modes of coregulation in
different brain tissues. In the probabilistic regulation programs
(PR), Hey2 is the highest ranked regulator for SC-10, consistently
assigned to the root level.  However, at level 1, there are three
equally good regulators Hes5 (overall rank 4), Neurod6 (overall rank
5) and Npas4 (overall rank 2).  For SC-15, Neurod6 is the highest
ranked regulator, consistently assigned to the root level, but the
assignment of Hey2 at level 1 has a very low score (overall rank 4).
For SC-27, we find consistent assignments of Hey2 at root level with
overall rank 1 and Neurod6 at level 1 with overall rank 2.  Thus the
cross-repression mechanism predicted by \cite{li2007} is supported
only in the case of SC-27 and not SC-10 and 15.  This example
underscores the usefulness of an ensemble method to assess confidence
levels of predicted interactions, especially in cases with limited
amount of expression data and no other validation sources available.

\section{Conclusions}

We have reexamined the module network method of \cite{segal2003} and
compared an ensemble-based strategy to the standard direct
optimization-based strategy. Ensemble averaging selects a subset of
most informative genes and builds a quantitatively better model for
them. It finds functionally more coherent tight gene clusters and is
able to determine the statistically most significant regulator
assignments. The difficult problem of identifying multiple regulators
which explain together, but not separately, the expression of a module
can be addressed in a reliable way.  The ensemble method is thus able
to deliver the promise to infer context-specific and combinatorial
regulation through the probabilistic module network model.

\section{Methods}
\label{sec:methods}

\subsection{Bayesian two-way clustering}
\label{sec:bayes-cocl}

We associate to each gene $i$ a continuous valued random variable
$X_i$ measuring the gene's expression level. For a data matrix
$\D=(x_{im})$ with expression values for $N$ genes in $M$ conditions,
the module network model of \cite{segal2003} gives rise to a
probabilistic model for two-way clusters, where a two-way cluster $k$
is defined as a subset of genes $\A_k\subset \{1,\dots,N\}$ with a
partition $\E_k$ of the set $\{1,\dots,M\}$ into condition clusters.
The Bayesian posterior probability for a set of coclusters
$(\A_k,\E_k)$, denoted $\C$, is given by
\begin{equation*}
  \ppost(\C) \propto \prod_k \prod_{E\in\E_k}
  \iint d\mu d\tau\; p(\mu,\tau) \prod_{i\in\A_k}\prod_{m\in E}
  p(x_{i,m}\mid \mu,\tau),
\end{equation*}
where $p(x\mid \mu,\tau)$ is a normal distribution with mean $\mu$ and
precision $\tau$ and $p(\mu,\tau)$ is a normal-gamma distribution (see
\cite{segal2005} or \cite{joshi2007} for more details). We use the
Gibbs sampler strategy developed in \cite{joshi2007} to sample
multiple high-scoring coclusterings from this posterior distribution.
From these multiple solutions we extract tight gene clusters using the
procedure outlined in \cite{joshi2007}. It consists of a graph
spectral method extracting densely connected regions from the graph on
the set of genes with edge-weights $p_{ij}$, the frequency that gene
$i$ and $j$ belong to the same cocluster in each of the sampled
solutions.

\subsection{Probabilistic regulatory programs}
\label{sec:prob-regul-progr}

For each set of conditions $E$ in the condition partition $\E_k$ for a
given module $k$ we have an associated normal distribution with
parameters $(\mu_E,\tau_E)$ which can be estimated from the posterior
distribution. Hence such a condition set can be interpreted as a
discrete expression state for the module. A regulatory program
`predicts' the expression state of any condition in terms of the
expression levels of a small set $\Rr_k$ of regulators, i.e., there is
a conditional distribution
\begin{align*}
  p\bigl(x_i \mid \{x_r, r\in\Rr_k\}\bigr) = p(x_i\mid \mu_E,
  \tau_E).
\end{align*}

The selection of an expression state is done by constructing a
decision tree with the states $E\in\E_k$ at the leaves.  To each
internal node $t$, we associate a regulator $r_t$ and split value
$z_t$. In \cite{segal2003}, the decision at the node is based on
the test $x_{r_t}\geq z_t$ or $x_{r_t}<z_t$. Here we extend this model
to allow \emph{fuzzy} decision trees. More precisely, we sort the
expression states $E\in\E_k$ by their mean $\mu_E$, and link this
ordered set hierarchically. Then we can associate to each internal
node a binary variable $y_t=\pm 1$, where $y_t=-1$ means `decrease
expression state' (go `left' in decision tree) and $y_t=+1$ means
`increase expression state' (go `right' in decision tree). Again we
also associate a regulator $r_t$ and split value $z_t$ to node $t$,
and a conditional probability
\begin{equation}\label{eq:1}
  p(y_t\mid x_{r_t},z_t,\beta_t) = \frac1{1+e^{-\beta_t y_t
      (x_{r_t}-z_t)}}.
\end{equation}
Given expression values $x_r$ for all $r\in\Rr_k$, we traverse the
decision tree in a probabilistic fashion, taking the decision
$y_t=\pm1$ at each node $t$ by tossing a biased coin with bias eq.
(\ref{eq:1}). The original model with \emph{hard} decision trees is
recovered if $\beta_t=\pm\infty$ for each node.

The conditional distribution or regulatory program now becomes a
normal mixture distribution
\begin{equation}\label{eq:5}
  p\bigl(x_i \mid \{x_r, r\in\Rr_k\}\bigr) = \sum_{E\in\E_k}
  \alpha_E\bigl(\{x_r, r\in\Rr_k\}\bigr)\,p(x_i\mid \mu_E, \tau_E)
\end{equation}
where
\begin{align*}
  \alpha_E\bigl(\{x_r, r\in\Rr_k\}\bigr) = \prod_t p(y_t\mid
  x_{r_t},z_t,\beta_t)
\end{align*}
with the values $y_t$ determined by the unique path through the
decision tree that ends at leaf $E$.

For a cocluster $(\A_k,\E_k)$ inferred from a data set $\D=(x_{im})$
by the method summarized in the previous section, we can derive a
posterior probability function for each regulator at each node $t$ as
follows. First note that each condition $m$ belongs to exactly one set
$E$ in $\A_k$ and hence determines a unique path through the decision
tree, or in other words a set of values $y_{t,m}$ at each node $t$.
Furthermore, each node $t$ has an associated condition set $E_t$
consisting of the union of all condition sets $E$ which can be reached
from node $t$. Hence we can define at each node a posterior
probability by
\begin{equation}\label{eq:2}
  \ppost[(r,z)] \propto \max_{\beta}\Bigl(\prod_{m\in E_t}
  p(y_{t,m}\mid x_{r,m},z,\beta)\Bigr),
\end{equation}
where for computational simplicity we maximize over $\beta$ instead of
marginalizing over a prior distribution. By allowing only a discrete
set of split values, eq. (\ref{eq:2}) becomes a discrete distribution
from which it is easy to sample. Typically, we consider as possible
split values z the expression values $x_{r,m}$ for $m\in E_t$, but
simpler schemes such as only allowing one or two split values can be
used to reduce computation time for large data sets.

The posterior probability eq. (\ref{eq:2}) measures how well the
expression values of a regulator `predict' the partition into two sets
of $E_t$ induced by the condition partition $\E_k$. We define the
average prediction probability of $(r,z)$ at node $t$ by the geometric
average
\begin{equation}\label{eq:3}
  p_t(r,z) = \Bigl(\prod_{m\in E_t} p(y_{t,m}\mid
  x_{r,m},z,\beta_{\max})\Bigr)^{1/|E_t|},
\end{equation}
where $\beta_{\max}$ is the maximizer in eq. (\ref{eq:2}).

\subsection{Regulator assignment score}

To assess the significance $Z_t(r)$ for assigning a regulator $r$ to a
node $t$ in a certain regulation program, we use the average
prediction probabilities (eq.  (\ref{eq:3})) and define:
\begin{equation}\label{eq:4}
  Z_t(r) = w_t \sum_z p_t(r,z).
\end{equation}
A typical choice for the weight factor $w_t$ is $w_t =
\frac{|E_t|}{M}$, expressing that we have more confidence in
assignments to nodes supported on more conditions. The sum $\sum_z$
runs over the discrete set of split values for regulator $r$ at node
$t$.  The overall significance $Z(r)$ for assigning a regulator $r$ to
a module is defined by summing eq. (\ref{eq:4}) over all nodes of all
regulation programs for that module:
\begin{align*}
  Z(r) = \sum_{T\in\T} \sum_{t\in T} Z_t(r).
\end{align*}

\subsection{Model evaluation}
\label{sec:model-evaluation}

For an experiment with expression levels $(x_1,\dots,x_N)$, we can
evaluate the probability distribution 
\begin{align*}
  p(x_1,\dots,x_N) = \prod_{i=1}^N p\bigl(x_i\mid \{x_r,
  r\in\mathcal{R}_{k(i)}\}\bigr),
\end{align*}
with $k(i)$ the module to which gene $i$ belongs and $\mathcal{R}_k$
the regulator set of module $k$, using the conditional distributions
\eqref{eq:5}.  We only consider genes for which the model makes actual
predicitions, i.e., genes belonging to clusters with a regulation
tree.  For the cross-validation experiment, we removed $10\%$ of the
conditions randomly from the total of 173 conditions.  We learned
module networks on the remaining $90\%$ data and repeated this
procedure 10 times.

\subsection{Data sets}

Yeast expression data for 2355 differentially expressed genes in 173
stress conditions, gene clusters, their regulators, split values and
regression trees were downloaded from the supplemental website of
\cite{segal2003} at
\href{http://robotics.stanford.edu/~erans/module_nets/}%
{http://robotics.stanford.edu/$\sim$erans/module\_nets/}.  MIPS
functional catagories were downloaded from
\href{ftp://ftpmips.gsf.de/catalogue/annotation_data}%
{ftp://ftpmips.gsf.de/catalogue/annotation\_data}.  For TC and SC we
calculated the $p$-value whether the overlap between a given cluster
and a given functional catagory is statistically significant.  We used
data on genome-wide binding and phylogenetically conserved motifs for
102 transcription factors from \cite{harbison2004}. For a given
transcription factor, only genes that were bound with high confidence
(significance level $\alpha = 0.005$) and showed motif conservation in
at least one other \textit{Saccharomyces} species (besides \textit{S.
  cerevisiae}) were considered true targets.  We also downloaded all
known regulator target interactions from the YEASTRACT database
\href{http://www.yeastract.com}{http://www.yeastract.com}.  We
calculated the $p$-value whether the overlap between a given cluster
and a given transcription factor target set is statistically
significant.

Mouse expression data by \cite{su2004} was downloaded from
\href{http://wombat.gnf.org}{http://wombat.gnf.org} and the data
selection and normalization was done as described in \cite{li2007}.


\section*{Acknowledgments}

We would like to acknowledge Eric Bonnet for useful discussions.  AJ
is supported by an Early-Stage Marie Curie Fellowship.  This work is
supported by IWT (SBO-BioFrame) and IUAP P6/25 (BioMaGNet).\medskip


\end{document}